\documentclass[superscriptaddress,amsmath,amssymb,amsfonts,aps,prb,twocolumn,10pt,showkeys]{revtex4-1}

\usepackage{graphicx}
\usepackage{dcolumn}
\usepackage{color}
\usepackage{bm}
\usepackage{hyperref}
\usepackage[table]{xcolor}

\begin{document}

\title{An {\it ab initio} approach to anisotropic alloying into the Si(001) surface}

\author{D.~V. Prodan}
\email{dmitrii.prodan@skoltech.ru}
\affiliation{Skolkovo Institute of Science and Technology, Moscow 121205, Russia}

\author{G.~V. Paradezhenko}
\affiliation{Skolkovo Institute of Science and Technology, Moscow 121205, Russia}

\author{D. Yudin}
\affiliation{Skolkovo Institute of Science and Technology, Moscow 121205, Russia}

\author{A.~A. Pervishko}
\email{a.pervishko@skoltech.ru}
\affiliation{Skolkovo Institute of Science and Technology, Moscow 121205, Russia}

\date{\today}

\begin{abstract}
Employing density functional theory calculations we explore initial stage of competitive alloying of co-deposited silver and indium atoms into a silicon surface. Particularly, we identify respective adsorption positions and activation barriers governing their diffusion on the dimer-reconstructed silicon surface. Further, we develop a growth model that properly describes diffusion mechanisms and silicon morphology with the account of silicon dimerization and the presence of C-type defects. Based on the surface kinetic Monte Carlo simulations we examine dynamics of bimetallic adsorption and elaborate on the temperature effects on the submonolayer growth of Ag-In alloy. A close inspection of adatom migration clearly indicates effective nucleation of Ag and In atoms, followed by the formation of orthogonal one-dimensional atomic chains. We show that the epitaxial bimetal growth might potentially lead to exotic ordering of adatoms in the form of anisotropic two-dimensional lattices via orthogonal oriented single-atom wide metal rows. We argue that this scenario becomes favorable provided above room temperature, while our numerical results are shown to be in agreement with experimental findings.
\end{abstract}

\maketitle

\section{Introduction} 
In the response to the current technological demands for advanced optical and electronic properties, profound studies of surface morphology along with crystal growth methods constitute a major part of materials science and engineering~\cite{Duke1996,Dahman2017,Harraz2014,Zhang2013,ref:hamada2017}. For the last decades processes of adatom aggregation on a silicon substrate have been extensively researched from experimental and theoretical perspective. Not in the last place this is motivated by the idea of handling the properties of a system by deposition of various materials and designing unusual atomic ordering on the semiconductor surfaces~\cite{Dabrowski2000,Weisbuch2014,Himpsel2001}. This is particularly true for group III elements and noble metals which tend to aggregate in the form of adatom chains on a silicon surface at submonolayer coverage~\cite{Pena2022,Braun2017,Lelay1983,Takeuchi2000,Dai2004,Li2001,Wang2002,Takeuchi2000,Dong2001}. Different aspects of the initial nucleation and atomic arrangement have been approached using numerous experimental techniques, including the scanning tunneling microscopy (STM)~\cite{Samsavar1989,Hashizume1990,Lin1993,Hornvonhoegen1995,Glueckstein1996,Wang2002,Li2001,Dong2001,Javorsky2009,Nogami1999}, photoemission spectroscopy~\cite{Samsavar1988,Matsuda1999,Matsuda2001,Aristov2010}, and X-ray photoelectron diffraction~\cite{Shivaprasad1995, Shimomura1998,Bunk1999,Suzuki2017} to name a few. 

Customarily, using kinetic Monte Carlo (KMC) simulations serves as a means to keep track of the atomistic processes on the surface~\cite{ref:introductionKMC,Battaile2008,Chatterjee2007}. Thanks to its computational efficiency the structural properties of relatively large systems can be reliably assessed at various temperatures and reaching timescales up to seconds. To get proper results within KMC methodology, one has to determine all possible events adatoms can participate in and corresponding binding energies that can be estimated using first-principles methods~\cite{ref:albao2005,ref:auagnanowires,Tokar2006,Kocan2009,Tokar2015,Syromyatnikov2021}. Following previous studies on adsorption and aggregation of metal adatoms on the silicon reconstructed surface, one can discern the formation of chain-like structures perpendicular to the silicon dimer rows and islands of adsorbate on the surface with distinctive contribution from C-defects that might act as nucleation centers \cite{Wang2002,Samsavar1989,Lin1993,Javorsky2009,Takeuchi2000,Radny2010,Dai2004,Dong2001,Evans1999, Li2001,Nogami1999,Hashizume1990,Kocan2006Ag,Pieczyrak2014,Huang2021,Huang2022}. To give a quantitative estimate, we herein explore the idea of using a thin alloying layer that provides a natural playground to modify the electronic properties of parent material~\cite{ref:MahlbergPtRu,Okamoto2004,Gross2009,Osiecki2012,Gruznev2017,Denisov2014,Matetskiy2015,Zhang2018,Hsu2018,Matetskiy2018,Yu2012}. 

Recently, it was shown that bimetal-silicon systems offer significant possibilities towards self-assembled growth of heterogeneous structures~\cite{ref:auagnanowires,ref:auagfilms,Jure2000,Magaud2000,Magaud2002,Sobotik2013,Puchalska2013,Gruznev2020,Gruznev2020InAu,ref:wong2017}. In particular, experimental studies of the Si(001) surface morphology with simultaneously deposited Ag and In atoms demonstrate the formation of unusual two-dimensional trellis-like self-organized structures affected by surface defects~\cite{ref:bimetal_exp}. In this Paper, we report on the effect of Ag-In co-deposition on a silicon surface using first-principles-assisted KMC simulations, where the growth process and related structural changes are included by respective energy barriers as determined from the density functional theory (DFT) calculations. Within the broad temperature range we explore adatom ordering on the surface that subsequently allows to check the experimentally proposed model of Ag-In preferential arrangement with orthogonal Ag-In wire formation~\cite{ref:bimetal_exp}. The rest of the Paper is organized as follows. Sect.~2 is devoted to a systematic description of utilized computational methods. The numerical results of a single atom adsorption are presented in Sect.~3. In Sect.~4, we discuss the co-deposition of atoms on a silicon surface in reference to experiments, providing the insights into bimetallic growth process on the semiconductor surface.

\section{Computational methods}

\subsection{Density functional theory}
We address atomic adsorption on a silicon surface using the density functional theory calculations as implemented in the Vienna {\it ab-initio} simulation package (VASP)~\cite{VASP1,VASP2,VASP3,VASP4}, where the interaction between ions and valence electrons is described by the projector augmented wave (PAW) method~\cite{PAW} and the many-electron interactions are introduced according to the Perdew-Burke-Ernzerhof (PBE)~\cite{PBE} exchange-correlation energy functional. Throughout all surface calculations we use a $2\times4\times1$ Gamma-centered $k$-point sampling grid and restrict the plane wave kinetic energy cut-off to 400~eV. We also set a convergence criteria for the self-consistent iteration process of the relative energies to be below 0.01~meV and interatomic forces to be smaller than 0.01~eV/\AA.

\begin{figure}
\centering
\includegraphics[width=0.8\linewidth]{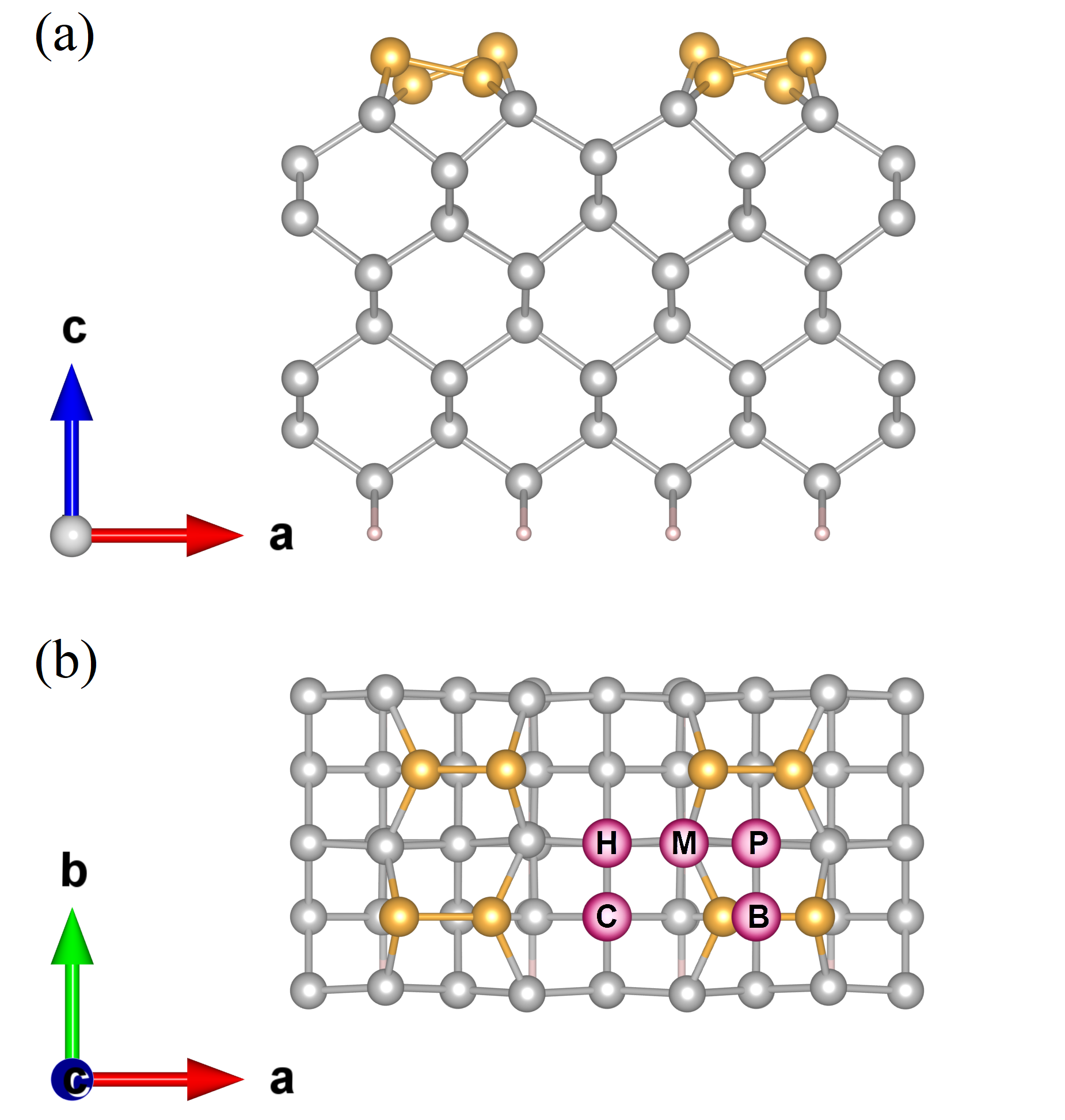}
\caption{Schematic side (a) and top (b) views of the clean reconstructed Si(001) surface with alternately buckled Si dimers. Buckled dimer atoms are colored by yellow, whereas gray atoms correspond to silicon in the layers below dimers. Possible adsorption sites for a single adatom on the reconstructed surface are labeled by P, C, B, H, and M on (b) representing pedestal, cave, bridge, hollow, and M sites, respectively.}
\label{fig:scheme}
\end{figure}

We model a clean silicon surface by a periodic slab of the finite thickness in the (001) direction that consists of eight silicon atomic layers accompanied by Si dimer reconstruction on one slab surface, while the other is passivated with hydrogen atoms (see Fig.~\ref{fig:scheme}). A vacuum of about 15~{\AA} is added on top of the considered structure to avoid the influence of periodic images of the slab. We take the $(4\times 2)$ surface supercell to address the possibility of surface reconstruction. During the slab relaxation, the atoms in three lower layers are kept frozen at the perfect crystal positions of bulk silicon with the lattice constant of 5.43~{\AA} (see e.g. Ref.~\cite{Ashman2004,Punkkinen2008,Fredrickson2016}), whereas the positions of the rest atoms are allowed to relax. 

To find the most favorable adsorption position of a single adatom we estimate the adsorption energy of an atom placed at selected number of sites as
\begin{equation}
E_{\rm ad}=E_{\rm at/Si}-E_{\rm Si}-E_{\rm at},
\label{eq:ad}
\end{equation}
where $E_{\rm at/Si}$ is the total energy of the silicon slab with atom located at chosen site and $E_{\rm Si}$ and $E_{\rm at}$ are the energies of clean silicon structure and free adatom, respectively.

The migration of adatoms on the Si(001) surface and related diffusion energy barriers that define 
the minimal energy path between various adatom binding sites are studied using the nudged elastic band (NEB) \cite{NEB} method. For post-processing of the calculated data the VASPKIT \cite{VASPKIT} tool is employed. In order to include the possibility of adatom chain formation, we construct $(8\times 4)$ surface supercell from the previously optimized geometry, where the number of silicon atomic layers is reduced to five, lower three atomic layers are fixed and match the positions in the relaxed $(4\times2)$ supercell. Since the NEB method is computationally heavier as compared to single relaxation procedure, we use a looser convergence threshold of 0.07~eV/{\AA}. We take six intermediate images to generate the initial diffusion path and set the spring force of constant 5~eV/{\AA}$^2$. The activation energy $E$ corresponding to the transition between two adatom sites is found as the difference between the energy of initial configuration and maximum energy along the diffusion path. 

\subsection{Kinetic Monte-Carlo}\label{sec:kmc}

As has been earlier discussed, the KMC approach has proved its efficiency in relatively large systems and over long timescales (up to seconds)~\cite{ref:bortz1975, ref:gillespie1978, ref:introductionKMC, ref:Hayakawa2021}. While the qualitative description of the conventional KMC can be found elsewhere, below we address specific amendments required for studying bimetallic growth of Ag and In adatoms on the $(2\times1)$-reconstructed Si(001) surface.

\begin{figure}[ht!]
\centering
\includegraphics[width=0.55\linewidth]{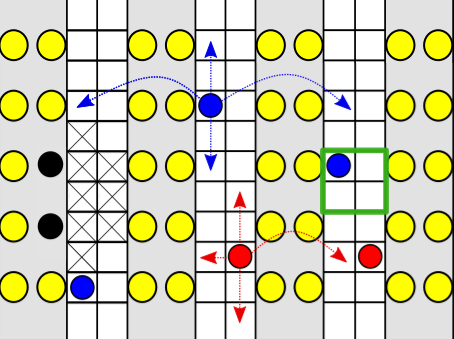}
\caption{Schematic of the lattice model. Yellow, blue, and red circles represent Si, Ag and In atoms, respectively, while black circles mark C-type defect position. The arrows mark possible hopping directions for Ag and In adatoms, while crosses represent forbidden lattice sites due to the presence of neighboring C-defect. The SDRs are depicted by gray. The simulation lattice is formed by $2\times2$ unit cells located between the SDRs. Single $2\times2$ unit cell is highlighted by green frame.}
\label{fig:adatoms_arrangement}
\end{figure}

In practice, process of adatom growth on the $(2\times1)$ reconstructed surface can be considered as a series of adatom hoppings on an anisotropic lattice (see Fig.~\ref{fig:adatoms_arrangement}) having complex structure since deposited metals of Ag and In enjoy different preferential adsorption positions~\cite{ref:bimetal_exp}. Results of first-principles simulations suggest that Ag adatoms tend to occupy a single site located strictly between the silicon dimers at low coverage. Meanwhile, for In adatoms the energetically preferential lattice sites are located between the silicon dimers rows (SDR) with the offset by half a period along the SDR direction. Without loss of generality, in the follow-up analysis we orient SDRs vertically as shown in Fig.~\ref{fig:adatoms_arrangement}. Practically, the simulation lattice is formed by $2\times 2$ unit cells each of which is constituted by four lattice sites, including a site for Ag adatom, associated with a preferential cave site, two sites for In adatoms that allow to include the effect of dimerization between the SDRs, and one dummy site that is forbidden to occupy by adatoms as highlighted in Fig.~\ref{fig:adatoms_arrangement}. To address physical effects in the system, we assume that only one type of adatoms can occupy a unit cell simultaneously and introduce C-type defects distributed randomly on the $(2\times1)$-reconstructed Si(001) surface. Depending on its position each C-defect creates a number of forbidden neighboring sites also marked in Fig.~\ref{fig:adatoms_arrangement}, while acting as a nucleation center for both Ag and In adatoms at remaining sites. 

Following one-metal growth simulation routine~\cite{ref:albao2009, ref:albao2010, ref:albao2015} we consider four processes where adatoms are involved, namely are deposition, hopping (or diffusion), nucleation, and detachment. 

{\bf Deposition.} In our modeling, deposition of new adatoms occurs with a constant rate of $0.2$~ML/s with respect to different adatom types that is comparable to those of other processes in the system, especially at high temperatures~\cite{ref:albao2009, ref:albao2010, ref:albao2015}. The deposition sites are selected randomly provided the uniform distribution. If a chosen site is occupied or forbidden due to the presence of neighboring adatoms or C-type defects, the adjacent sites are checked. Adatom is placed at any free neighboring site, subject to availability; and deposition is rejected otherwise.

{\bf Hopping.} The hopping rates for deposited particles are governed by the Arrhenius law,
\begin{equation}\label{Arrhenius}
    R = \nu_0 \exp{\left(-\frac{E}{k_{\mathrm{B}} T} \right)},
\end{equation}
where $\nu_0$ is the hopping frequency set to be $10^{13}$~s$^{-1}$, $E$ is the activation energy, $k_{\rm B}$ is the Boltzmann constant, and $T$ is the temperature. The hopping of Ag and In adatoms can occur either parallel or perpendicular to the SDRs direction. 

{\bf Nucleation.} In the model, we account for four possible scenarios of adatoms nucleation resulting in single-species In and Ag adatom islands, combined Ag-In structures, as well as patterns formed due to the presence of C-type defects. Particularly, it is assumed that In island emerges when two adatoms meet in adjacent sites attributed to the same unit cell, while two Ag adatoms form an island when they occur in neighbouring unit cells~\cite{ref:bimetal_exp}. A pair of In and Ag adatoms nucleates if they appear in lattice sites located in adjacent unit cells separated by SDR. Additionally, C-type defects can capture adatoms, leading to In and Ag adatom nucleation when they end up in the allowed lattice sites in the vicinity of C-defect.

{\bf Detachment.} We also introduce a process of a single adatom detachment from a nucleated island that similarly to the hopping rate is described by the Arrhenius law~\eqref{Arrhenius}. The activation energy for a detachment process depends on the adatom type, hopping direction, and \emph{local} environment of neighboring adatoms and C-defects. When detachment occurs,  selected adatom breaks away from the island and travels to available adjacent lattice site in the chosen direction. 

\begin{figure}[ht!]
\includegraphics[width=1.0\linewidth]{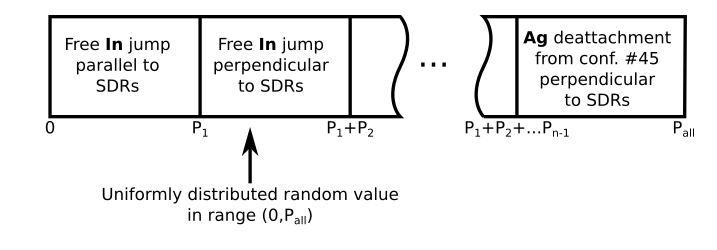}
\caption{Scheme of choosing the next type event in the KMC algorithm.}
\label{fig:KCM_scheme}
\end{figure}

In practice, the KMC algorithm to feature bimetal growth is organized as follows. By convention, $R_1$ is the hopping rate of a free In adatom in the direction parallel to the SDRs, $R_2$ is the hopping rate of a free In adatom in the direction perpendicular to the SDRs, etc. For each rate $R_i$ we store and dynamically update 
the list of $N_i$ adatoms acceptable for this type of process. Events are chosen according to the Gillespie scheme~\cite{ref:gillespie1977}. On each step of the algorithm, we calculate the probabilities of every specific process $P_i = R_i N_i$ and the total probability $P_{\rm all} = \sum_i P_i$. Then, we generate a random number and select the next type of event based on these probabilities as shown in Fig.~\ref{fig:KCM_scheme}. We randomly choose one of $N_i$ adatoms from the corresponding list and update its position according to the chosen type of event. Finally, we update the lists of events and recalculate the probabilities $P_i$ and $P_{\rm all}$.

\section{Adsorption of a single atom}
Before we study preferential adatom positions on a silicon surface within first-principles approach, we first perform clean Si(001) slab optimization and obtain that the Si(001) surface undergoes $c(4\times 2)$ reconstruction favoring the formation of alternating buckled dimers with bond length of 2.35~{\AA} over the other reconstructions, including the symmetric case with dimer bonds parallel to the surface, that agrees well with previous theoretical studies~\cite{Chadi1979,Roberts1990,Brocks1991,Dabrowski1992,Northrup1993,Pehlke1993,Cho1994,Ramstad1995,Stekolnikov2002,Garrity2009,Czekala2014,Guo2014} and experimental findings~\cite{Wertheim1991,Wolkow1992,Tochihara1994,Shirasawa2006,Back2013}.

To find plausible adsorption sites for indium and silver atoms to occupy, we consider five non-equivalent positions: pedestal, cave, bridge, hollow, and M, shown schematically in Fig.~\ref{fig:scheme}b. We locate pedestal and cave sites at the center between neighboring silicon dimers within the same and adjacent SDRs, respectively. Bridge point is positioned above the middle of the silicon dimer bond. Hollow and M sites are between two neighbouring SDRs, where the former is equidistant from the nearest silicon atoms composing dimers and the later is off-center binding site.

We analyse preferential adsorption sites by calculating the adsorption energy per single adatom as given by Eq.~\ref{eq:ad}. The respective adsorption energies for Ag and In atoms are summarized in Table~\ref{tab:ad}. The cave site is found to be the preferred adsorption site over considered for single Ag atom~\cite{Zhou1993,Kong2003,Huxter2019}, while the M site is the most favorable adsorption position for In atom being slightly more energetically stable than the cave site~\cite{Takeuchi2000,ref:albao2009,ref:albao2010}. It should also be mentioned that indium atoms are known to grow in one-dimensional dimer chains perpendicularly to the underlying SDRs, which, in turn, is more energetically stable than attachment of a single atom~\cite{Northrup1991,Dong1997,Evans1999,Takeuchi2000,Dai2004,Kocan2008,Kim2009}. Therefore, in subsequent KMC simulations we expect occupation of silver atoms between adjacent SDRs on the line connecting neighboring dimer bonds that follows the experiments on Ag growth on the Si(001) surface highlighting the effect of isolated atom adsorption at low coverage~\cite{Samsavar1989,Lin1993,Shivaprasad1995,Cho1999,Huang2021} that later than provides the path for further Ag islands formation~\cite{Zhou1993,Kocan2006Ag}. While being out of scope of the current study it has to be stressed that coverage down to a monolayer limit results in more complex interactions between Ag and Si leading to dimer and tetramer silver formation~\cite{Kong2003,Huxter2019}. For indium we restrict individual atom to reside at the M site and, in addition, allow second In atom to join it causing the dimer to appear~\cite{Evans1999,Kocan2008}.

\begin{table}[ht!]
\small
\caption{Calculated adsorption energies for considered sites (see Fig.~\ref{fig:scheme}), given in eV.\label{tab:ad}}
\begin{tabular*}{0.48\textwidth}{l@{\extracolsep{\fill}}llllll}
\hline
&pedestal&cave&bridge&hollow&M \\
\hline
\hline
Ag&-2.435&-2.774&-2.205&-1.953&-2.405\\
In&-2.494&-2.713&-2.518&-2.653&-2.715\\ 
\hline
\end{tabular*}
\end{table}

Preparation conditions of the clean Si(001) samples are inextricably linked to surface defect formation. Conventionally, one distinguishes three types of defects on Si(001), named A, B, and C-defects, where the former two are vacancy defects interpreted as a missing single and a pair of silicon dimers at the surface, respectively. Whereas largely debated C-defect is attributed to dissociative adsorption of a single water molecule on two adjacent dimers~\cite{Hamers1989,Hossain2003,Okano2004,Tanaka2008,Pieczyrak2014}. The later surface impurity is commonly encountered in STM measurements~\cite{Hata2000,Nishizawa2002,Warschkow2008,Yu2008,Sobotik2008,Yu2011} and was found to play an important role at adsorption of In \cite{Kocan2007,Javorsky2009,Kocan2008,Ostadal2008} and Ag \cite{Kocan2006Ag,Pieczyrak2014} adatoms diffusing on the silicon surface being preferential pinning position for adatom chain formation. For this reason, C-type defects should also be taken into account during the analysis of different adsorption geometries, while the other defects that are not nucleation centers can be disregarded for simplicity. Similarly to previous studies, in our DFT calculations, we introduce the C-defect by hydrogen atom and hydroxyl group bonded to neighboring silicon atoms attributed to adjacent dimers within the same SDR.

\section{Co-deposited atom diffusion}
With the preferential adsorption sites for a single adatom on the silicon surface being identified, it is possible to search for the minimum energy pathways connecting two nearest adsorption minima for considered {\it{local}} adatom arrangements, and calculate the transition barrier between these states. We employ the NEB method to compute migration pathways starting from the preferential adatom position in the selected optimized configuration to the nearest allowed binding site. 

We obtain that migration of a single Ag atom on Si(001) is highly anisotropic, favoring the adatom motion in the direction perpendicular to the SDRs that agrees well with previously reported data~\cite{Zhou1993}. In contrast, diffusion barriers determined from the simulations for a single In adatom are almost isotropic~\cite{Javorsky2009}. Interestingly, for both types of atoms configurations of a single atom with C-type defect are energetically more stable which might trigger adatom nucleation process and subsequent island growth on the surface within KMC modeling. It is worth mentioning that calculated transitions along the minimum energy path result in the silicon dimer flipping and contributing to adatom dynamics. 

\begin{figure*}[t]
    \centering
    \includegraphics[width=0.9\linewidth]{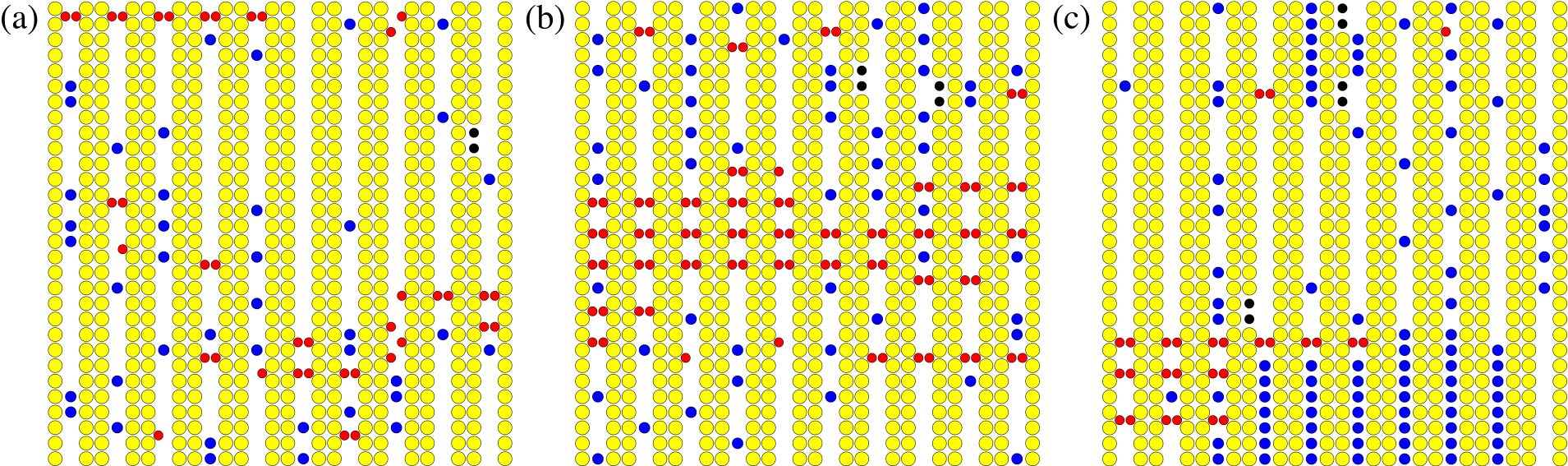}
    \caption{Snapshots of Ag-In adatom arrangements obtained by the KMC simulations at final coverage $\theta=0.1$~ML for different temperatures: (a) $T=150$~K, (b) $T=300$~K, and (c) $T=500$~K. The red and blue circles depict In and Ag adatoms, respectively. The silicon atoms forming SDRs are introduced by yellow circles.}
    \label{fig:visualisations}
\end{figure*}

Following the discussed routine in Sect.~\ref{sec:kmc} and results of first-principles modeling, we perform the KMC simulations on a $100\times201$ lattice in the temperature range from 150~K to 500~K. The final coverage is set 0.1~ML. The fragments of simulated surface configurations at different temperatures are presented in Fig.~\ref{fig:visualisations}. The KMC findings highlight the crucial role of In adatoms in orthogonal oriented bimetal row growth. As one can see in Fig.~\ref{fig:visualisations}, In adatoms have a tendency to assembly in the chains of dimers aligned perpendicular to the SDRs. At low temperatures, we observe a plenty of short In chains (up to 10~datoms) as well as separate In dimers and monomers, whereas at high temperatures, In adatoms arrange in longer chains of several dozens of indium atoms and single dimer/monomer configurations appear rarely. Meanwhile, silver atoms are not involved in island formation at low temperatures being generally fixed at preferential deposition site (see, e.g., Fig.~\ref{fig:visualisations}a at $T=150$ K). As the temperature increases, Ag adatoms start to diffuse on the silicon surface and form chains attached to In islands in the direction parallel to the SDRs. Particularly, at $T=300$ K (Fig.~\ref{fig:visualisations}b) silver chains are quite rare and short (up to 5 adatoms), but their number and length increase with temperature resulting in a crosswise bimetal adatom structure, where In and Ag chains are oriented perpendicular and parallel to the SDRs, respectively. The simulation results qualitatively agree with the STM images reported at room temperature and at 100~$^\circ$C in Ref.~\cite{ref:bimetal_exp}. However, it should be stressed that in our simulations parallel Ag chains appear regularly and separated by one SDR, while in the reported STM images~\cite{ref:bimetal_exp} chains attributed to Ag adatoms occur on a larger scale of several SDRs. At the same time, the C-defects have implicit effect in adatom growth process. We obtain that C-type defects become preferential pinning position largely for Ag adatoms, while islands of In chains appear regardless of C-defects. Nevertheless, as it was found during the modeling in the absence of C-type defects, In adatoms migrate along the lattice more often affecting island nucleation process.

\begin{figure}
    \centering
    \includegraphics[width=0.8\linewidth]{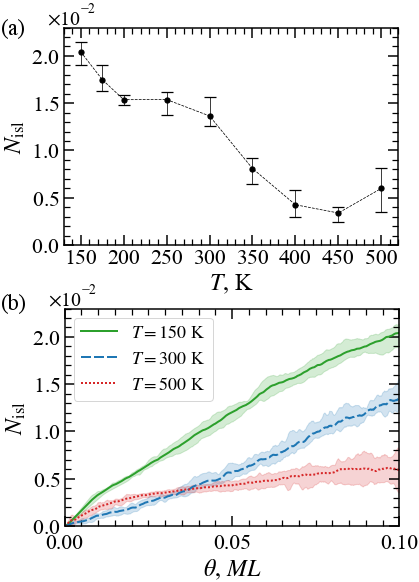}
    \caption{Mean island density $N_{\rm isl}$ calculated within the KMC approach depending on (a) temperature $T$, where the coverage is fixed at $\theta=0.1$~ML; and (b) coverage $\theta$ for selected temperatures $T=150$, $300$, and $500$~K. Presented results are averaged over several simulation runs.}
    \label{fig:n_t}
\end{figure}

We also perform a series of simulation runs at different temperatures to calculate the mean island density providing quantitative description of two-dimensional structure growth process and given by~\cite{ref:albao2005} 
\begin{equation}\label{N-isl}
    N_{\rm isl}=\frac{\theta}{\langle s \rangle},
\end{equation}
where $\theta$ is the surface coverage, and $\langle s \rangle$ is the mean island size measured in the number of adatoms.

First, we investigate the dependence of mean island density~\eqref{N-isl} on temperature at fixed final coverage $\theta = 0.1$~ML (Fig.~\ref{fig:n_t}a). One can notice a general trend of $N_{\rm isl}$ to decrease with temperature that follows respected atomic arrangements shown in Fig.~\ref{fig:visualisations}. At low temperatures, surface configurations are mostly represented by small islands and single adatoms resulting in a relatively small mean island size~$\langle s \rangle$ (Fig.~\ref{fig:visualisations}a) and subsequently high mean island density. As temperature increases  (Figs.~\ref{fig:visualisations}bc), adatoms start to assemble in larger ordered structures leading to $N_{\rm isl}$ being decreased. Note that the same behaviour has been observed experimentally~\cite{ref:bimetal_exp}. Next, we study the mean island density as a function of coverage $\theta$ until the latter reaches its final threshold at $0.1$ ML (Fig.~\ref{fig:n_t}b).
At $T=150$~K and $300$~K, $N_{\rm isl}$ increases almost linearly with $\theta$ due to low adatom mobility that prevents adatom from diffusing and keeps the mean island size relatively constant with $\theta$. For higher temperatures, the slope of $N_{\rm isl}$ modifies under large-scale ordered structure growth. Interestingly, at specific $\theta$ new adatom islands stop to develop in the system, while the mean island size increases simultaneously with coverage. 

\section{Conclusions}
We provided a detailed theoretical study of anisotropic alloying leading to bimetal nanostructure formation on a silicon surface. Using developed model that takes into account various effects appearing on the surface, including adatom diffusion and detachment, we explored nucleation patterns caused by substrate features as well as interactions between adatoms in the system. Our results are in a good agreement with STM measurements. Particularly, we discovered mutual stabilization pattern between In and Ag adatoms and elaborated on the contribution of C-type surface defects on adatom nanostructures stability. The developed approach validates anisotropic two-dimensional lattice formation on the $(2\times1)$-reconstructed Si(001) surface depending on preparation conditions, and might be of potential interest in silicon-based technology.

\acknowledgements
We acknowledge use of the computational resources at the Skoltech supercomputer ``Zhores''~\cite{Zhores} in our numerical simulations. The work of A.A.P. was supported by the Russian Science Foundation Project No. 22-72-00021. D.Y. acknowledges the support from the Russian Science Foundation Project No. 22-11-00074.

\bibliographystyle{apsrev4-2}
\bibliography{text.bbl}

\end{document}